\documentclass[twocolumn]{aa}
\usepackage{graphics}
\input{epsf}
\newcommand{\bfo}[1]{\mbox{\boldmath $#1$}}
\def\bvarphi{\mbox{\boldmath $\varphi$}}

\begin{document}
\newcommand{\beq}{\begin{equation}}
\newcommand{\eeq}{\end{equation}}
%****** Begin Definitions *************************
\def\la{\hbox{\raise.35ex\rlap{$<$}\lower.6ex\hbox{$\sim$}\ }}
\def\ga{\hbox{\raise.35ex\rlap{$>$}\lower.6ex\hbox{$\sim$}\ }}
\def\runit{\hat {\bf  r}}
\def\phunit{\hat {\bfo \bvarphi}}
\def\zunit{\hat {\bf z}}
\def\beq{\begin{equation}}
\def\eeq{\end{equation}}
\def\beqa{\begin{eqnarray}}
\def\eeqa{\end{eqnarray}}
\def\sub#1{_{_{#1}}}
\def\order#1{{\cal O}\left({#1}\right)}
\newcommand{\sfrac}[2]{\small \mbox{$\frac{#1}{#2}$}}
%****** End Definitions ***************************
\title{On the nature of the hydrodynamic stability of accretion disks
%The compressible inviscid oscillating algebraic instability for
%azimuthally independent disturbances in thin circumstellar disks
    \thanks{Research supported in part by
    the Israel Science Foundation, the Helen and Robert Asher
    Fund and the Technion Fund for the Promotion of Research}}

\author{  O.M. Umurhan \& G. Shaviv}

\offprints{O.M. Umurhan, \email{mumurhan@physics.technion.ac.il}}

\institute{Department of Physics, Technion-Israel Institute of
Technology,
  32000 Haifa, Israel}

\date{Received ---- / Accepted ----}
\titlerunning{Algebraic Growth in Circumstellar Disks}

\abstract{The linear stability of accretion disks is revisited.
The governing equations are expanded asymptotically and solved to first
order in the expansion parameter $\epsilon$ defined by the ratio of the
disk's vertical thickness to its radial extent. Algebraically growing solutions are found for global perturbations on the radial accretion flow of thin inviscid compressible
Keplerian disks.  The algebraic temporal behavior is exhibited in the vertical velocities
and the thermodynamic variables and has the form $t\sin\Omega_0 t$ locally in the
disk where $\Omega_0$ is the Keplerian rotation rate.  The physical implications and relations
to the Solberg-Hoiland stability criteria
are discussed.
\keywords{hydrodynamics -- accretion disks -- transient growth }}
  \maketitle
\section{Introduction}
It is well known that the pure Keplerian shear flow (KSF in this letter) is  linearly stable,
i.e. no exponentially growing solutions are known.
This stability analysis is implemented in accretion disks (AD) to also
imply their linear stability.
However, the governing equations for AD entail the existence of meridional circulation in the disk
(Kluzniak \& Kita, 2000, Regev \& Gitelman, 2002)
and here we are interested in how weak meridional flow affects the linear
stability of ADs.

A classical approximation in ADs is to integrate over the thin vertical extent of the disk.
However, since the dominant energy transfer is in this direction, this may result in a loss of some
salient physical features. Furthermore, it is also known that the meridional
circulation has a significant component in this
direction. Hence, the present work includes the vertical dimension explicitly.

To the point: the possibility that transient growth (TG) phenomena may play a significant role in
astrophysical disks has gained recent attention with the work
of Iaonnou and Kakouris (2001) and Yecko (2004).
The essence of the TG phenomenon is that
significant growth of an initial perturbation develops (up to orders of magnitude) before final decay.
(cf.  for a review see Schmid \& Henningson, 2001).
Indeed, linear stability analysis of  KSFs
show that asymptotic stability is promoted because
(i) the Rossby number of the flow is $\sim  1 $  and, (ii) there are no
inflection points in the KSF profile (Balbus, 2003), except in the star-disk
boundary layer (Bertout \& Regev, 1996).

Even though linear disturbances decay in the long run, the fate
and character of short time transient responses (and its nonlinear consequences)
in an AD may be significant and deserves deeper exploration.
Two observations provide circumstantial support to the idea of TG
of KSFs
: (a) the analogy to laboratory flows
like Couette-Taylor experiments (Richard \& Zahn, 1999,  Longaretti, 2002)
which have many common  features with KSFs and (b) linear stability
analysis of shearing flows in localized sections of KSFs
show significant TG (Chagelelishvili et al, 2003,
Tevzadze et al, 2003, Yecko, 2004, Mukhopadhyay et al., 2004, Afshordi et al., 2004) leading to significant nonlinear ramifications
(e.g. 2D simulations of Umurhan \& Regev, 2004).  Also,
Iaonnou \& Kakouris (2001) demonstrate TG for global disk perturbations
but their considerations were restricted to two-dimensional (radial-azimuthal)
disturbances of an incompressible KSF.
Studies of viscous compressible boundary layer shear flows
have been shown by Hanifi \& Henningson (1998) to
have solutions which grow like $t$ in the inviscid limit.
%They showed
%that these solutions are the counterparts to the solutions derived
%by Ellingson \& Palm (1975) for the problem of incompressible inviscid flow.
What
is most fascinating is that these algebraically growing solutions
are present in the limit where the streamwise perturbations are absent.
%and,
%they also demonstrate that it is on the back's of these solutions that the TG
%phenomenon observed in studies of viscous compressible flows are inextricably built.
These trends beget the question: is there some sort of analogous compressible
TG effect in accretion disks in the inviscid limit?
\par
In this work we show that algebraically growing solutions indeed do exist for global perturbations
of a geometrically thin AD.  The steady state is assumed to be a general barotrope and we consider relatively fast  adiabatic perturbations.
The equations studied here emerge from an asymptotic analysis of the equations appropriate for
such a disk environment (Regev, 1983, Kluzniak \& Kita, 2000,
Regev \& Gitelman, 2002)
%and, most recently, the dynamical expansions detailed in Umurhan et al. 2005),
in the zero viscous limit.

 \section{Equations and general solution}
 We assume that what governs the evolution of the  geometrically thin
 ADs are the Euler equations (inviscid flow).  By 'geometrically thin'
 we mean that
  \beqa
 \epsilon \equiv  {\tilde H}/   {\tilde R} \ll 1,
\eeqa
where the vertical and radial scales are $\tilde H$ and
$\tilde R$ respectively.
The governing equations are asymptotically expanded in  $\epsilon$ and
this is detailed in Appendix A.  We assume the expansion converges fast for sufficiently small $\epsilon$.
The physical advantage of the asymptotic expansion is
that the procedure draws out of the governing equations terms with
equal physical importance, namely the terms with the same $\epsilon$ power have the same order of magnitude effect on the flow.
Further, we assume here that the perturbations can be  treated adiabatically, namely we restrict the discussion to time scales shorter than the dissipation timescales in the AD.
 The three momentum conservation equations governing the evolution of disturbances
 in cylindrical coordinates  are (to the lowest non vanishing power in $\epsilon$),
 \beqa
 \partial_t u_1' &=& 2\Omega_0r\Omega_2' \label{eq:u} \\
 r\partial_t \Omega_2' &=& - u_1' \sfrac{1}{r}\partial_r r^2 \Omega_0,
\label{eq:Omega} \\
\rho_0{\partial_t v_2'} &=& - \partial_z P_2' - \rho_2' g(r,z),
\label{eq:v}
 \eeqa
where the dynamical radial ($r$), azimuthal ($\phi$) and vertical ($z$) velocities
are denoted by $u_1',\Omega_2',v_2'$ respectively.
The mass continuity and
energy conservation equations are
 \beqa
\partial_t \rho_2' &=& -\sfrac{1}{r}{\partial_r} \rho_0 r u_1'
 -{\partial_z}\rho_0 v_2', \\
\partial_t P_2' &=& -u_1'\partial_r P_0
-v_2'\partial_z P_0 - \gamma P_0\left(\sfrac{1}{r}{\partial_r r u_1'}
+{\partial_z v_2'}
\right).
\label{eq:P}
 \eeqa
The above are nondimensionalized according to the procedure described in Appendix A.
We stress that the undisturbed state contains the possibility of {\em steady} meridional circulation
only for non-zero viscosities.
The pressure and density disturbances are $P_2',\rho_2'$ while their steady
state configurations are $P_0,\rho_0$.  The vertical component of
gravity (which points towards $z=0$) is expressed generally as a function of $r$ and $z$;
for thin  rotationally supported Keplerian disks it is given by, $
g(z,r) = {z}/{r^3}$,
up to lowest nontrivial order in the $\epsilon$ expansion.
The steady rotation rate for Keplerian flow is given as $\Omega_0 = r^{-3/2}$.  The ratio of
specific heats is $\gamma$.\par

In the spirit of this general
discussion, it is assumed that the steady state thermodynamic
quantities behave as barotropes which means that there
is a unique function relating $P_0 = P_0(\rho_0)$.
%\beq P_0 =
%P_0(\rho_0), \qquad \frac{1}{n+1} \equiv 1 -
%\frac{P_0}{\rho_0}\frac{d\rho_0}{d P_0}.
%\eeq
For general barotropes the spatially dependent {\em
barotropic index}, $n(r,z)$, is given by $\frac{n+1}{n} \equiv \frac{d\ln P_0}{d\ln\rho_0}$.
The steady state relationship satisfies the vertical hydrostatic
equilibrium, namely,
$ {\partial_z P_0} = -g(z,r)\rho_0. $
It should be noted that integration of this
equation introduces an arbitrary function of the variable
$r$ which is usually taken to be the {\em height} of the disk, $h(r)$, which
is assumed to be well behaved.
%\footnote{The arbitrary nature of $h$ arises because of our neglect
%of other physical processes like turbulent viscosity or some kind of cooling effects or radiation.
%Since this is a general discussion, we only presume that $h$ never
%breaks its ordering.}
Finally, we define the adiabatic sound speed as
$c_0^2 \equiv \gamma P_0/\rho_0$. Once the form of the barotrope is known
then so is the sound speed.
Eqs.(\ref{eq:u}-\ref{eq:Omega}) may be reduced to the simple
PDE,
\beqa
\left(\partial_t^2 + \Omega_e^2\right)u_1' = 0,
\quad {\jmath} \equiv r^2\Omega_0, \quad
\Omega_e^2 = \sfrac{1}{r^3}\partial_r {\jmath}^2.
\label{combined:u:eq}
\eeqa
If as an initial condition we have an arbitrary meridional flow field and zero
azimuthal flow perturbation, i.e. at $t=0$, $u_1'=\bar u(r,z),
\ \ \Omega_2'=0$, then the solution
to (\ref{combined:u:eq}) is
\beq
u_1' = \bar u (r,z) \cos\Omega_e t, \qquad
 \Omega_2' = \bar\Omega \sin\Omega_e t,
\label{uprime_sol}
\eeq
with $\bar \Omega = -{\bar u}(r,z)/{2r^{1/2}}$.
The {\em epicyclic frequency} $\Omega_e$ is equal to $\Omega_0$
in KSF's.
Note that because $\Omega_e$ is
a function of $r$, inspection of the solution at this order
reveals its inseperability.
Combining (\ref{eq:v}-\ref{eq:P}) reveals the single PDE,
\beqa
\left({\partial_t^2} + {\cal L}\right)\rho_0 v_2' &=&
{\partial_z}\left(u_1'{\partial_r P_0}
+\sfrac{\gamma P_0}{r}{\partial_r ru_1'}\right)
+\sfrac{g}{r}{\partial_r \rho_0ru_1'}, \ \ \
\label{acoustic:eq}
\eeqa
where the differential operator ${\cal L}$ is given as
\beq
\left[\partial_z\left(g\sfrac{n+1 -n\gamma}{n+1}\right)\right]- \sfrac{n-1}{n+1}\gamma g{\partial_z}
- c_0^2{\partial_z^2}.
%\frac{\partial}{\partial z}\left[g
%\left(1 - \frac{n\gamma}{n+1}\right)\right]- \frac{n-1}{n+1}\gamma g\frac{\partial}{\partial z}
%- c_0^2\frac{\partial^2}{\partial z^2}.
\eeq
%The subscript $z$ above means partial $z$-differentiation of the quantity in the brackets.
We notice immediately that the evolution of $v_2'$ is explicitly driven by the solution obtained for $u_1'$.
With $\rho_2'=P_2' = 0$ at $t=0$ we have the solutions,
%\begin{mathletters}
\begin{eqnarray}
\rho_0 v_2' &=& M_v^{(1)}t\sin\Omega_e t + M_v^{(0)}\cos\Omega_e t, \nonumber\\
\rho_2' &=& \rho^{(1)}t\cos\Omega_e t + \rho^{(0)}\sin\Omega_e t, \nonumber\\
P_2' &=& P^{(1)}t\cos\Omega_e t + P^{(0)}\sin\Omega_e t.
\end{eqnarray}
The structure vector ${\bf M}= (M_v^{(1)},M_v^{(0)})$ satisfies
\begin{equation}
\label{eq:struct}
\left[{\cal L} - \Omega_e^2\right]M_v^{(i)} = \Phi_{i}~~~ i=0,1
\end{equation}
where the source vector $\Phi$,  the driver, is given by:
\begin{equation}
\Phi = \left\{ -\phi_1,  -\left(\phi_0 + 2\Omega_e M_v^{(1)}\right)\right\},
\end{equation}
in which
\beqa
\phi_0 &=& {\partial_z}\left(
\bar u {\partial_r P_0} \right)
+  {\partial_z}\left(
\gamma P_0\sfrac{1}{r} \partial_r r\bar u \right)
+\sfrac{g}{r}{\partial_r}\rho_0 r \bar u,
\nonumber \\
\phi_1 &=& -
\left({\partial_z}\gamma P_0 \bar u + g \rho_0 \bar u\right)
\partial_r\Omega_e.
\eeqa
The remaining structure functions are given to be
\beqa
\rho^{(1)} &=& \sfrac{1}{\Omega_e}\left(
\partial_z M_v^{(1)} - \rho_0 \bar u{\partial_r \Omega_e}
\right)
, \nonumber \\
\rho^{(0)} &=& -\sfrac{1}{\Omega_e}
\left(\sfrac{1}{r}\partial_r r\rho_0\bar u
+\partial_z M_v^{(0)} +\rho^{(1)}
\right), \nonumber \\
P^{(1)} &=&-\sfrac{1}{\Omega_e}\left[
gM_v^{(1)} - \gamma P_0
{\partial_z}({M_v^{(1)}}/{\rho_0}
+\gamma P_0 \bar u {\partial_r \Omega_e}\right],
\nonumber \\
\Omega_e P^{(0)} &=&
-P^{(1)} - \bar u\partial_r P_0 -
\gamma P_0\left(\sfrac{1}{r}\partial_r r\bar u +
\partial_z\sfrac{{M_v^{(0)}}}{\rho_0}\right)
+gM_v^{(0)}
\nonumber
%P^{(0)} &=&
%\frac{1}{\Omega_e}\Biggl[
%-P^{(1)} - \bar u \frac{\partial P_0}{\partial r} -
%\gamma P_0\frac{1}{r}\frac{\partial r\bar u}{\partial r} +  \nonumber \\
%& & \ \ \ \ \ \ \ \ \ \ \ \ \
%gM_v^{(0)} - \gamma P_0\frac{\partial}{\partial z}\left(\frac{M_v^{(0)}}{\rho_0}\right)
%\Biggr ].
\eeqa
For ${\partial_r \Omega_e} \neq 0$
the expression for $\phi_1$ is in general not zero.
%\footnote{
%{{\it Since the solution $u'_1$ is inseperable, radial derivatives of
%it bring out mulitplicative factors of $t$, e.g.
%${\partial u'_1\over \partial r} =
% t{\partial\Omega_e \over \partial r}\sin \Omega_e t$.
%Because the pde equation
%for $v'_2$
%has for source terms various derivative expressions
%of $u'_1$
%is
%why the solutions for the vertical velocity and the thermodynamic
%quantities have this algebraic time dependence.
%}
%}}
The above family of solutions  is completely determined
once the initial perturbation in the radial (meridional)
velocity, $\bar u$ (or equivalently ${\bar {\dot m}}=\rho_0 {\bar u}$), is given
(in contrast to a density or a pressure perturbation).  Next, the structure vector ${\bf M}$
must be determined subject to the boundary conditions.
The classical assumption is the  vanishing of the Lagrangian pressure perturbation   on the
moving surface at $z_s = \pm h(r,t)$
(e.g.  Korycansky \& Pringle, 1995).
We also study the homogeneous solutions
of (\ref{acoustic:eq}).  These acoustic modes
(i) oscillate stably and (ii) have
no frequencies  which are resonant with $\sin\Omega_e t$ (for $n$ constant).
%\begin{equation}
%\left(
%{\cal L} + \frac{\partial^2}{\partial t^2}\right)\rho_0 v = 0,
%\label{homo_acoustics}
%\end{equation}
\section{Nature of the solution}
Classical  stability analysis of linear systems assumes solutions of the simple $\exp(i \omega t)$ form and searches for a dispersion relation for $\omega$. Here we were able to find a complete solution
(i.e. $\sim t\sin\Omega_e t$)
that cannot be represented by a {\it finite} number of exponential functions. The basic reason for this behavior lies in the structure of the properties  of the $\epsilon$ expansion as reflected in the
the set (\ref{combined:u:eq}-\ref{acoustic:eq}) and revealed in
(\ref{eq:struct}). The equation for the first component is linear in the perturbation. The second equation is also linear however, the second equation has as  a source the solution to the first equation. This is a system of staggered linear equations where at each level the solutions to the previous level appear as source terms.

%Overlooking the physical origin of the system of equations governing a vertically thin structure,
%the mathematical roots for this unique solution can be traced to the structure of the driver vector.
%The first component does not depend on the second component while the second component depends on the first one.
%The shear (i.e. $d\Omega_e/dr$) appears only in the first term.
%We find that $\phi_1$ is composed of the shear and the structure form in the vertical direction only.
%The second term contains the resulting structure vector plus a small correction $\phi_0$ from the radial
%structure. The operator ${\cal L}$ depends solely on the vertical coordinate.

The physical effect responsible for the algebraic growth can be traced from the
driver vector $\Phi$ to arise from the basic shear, i.e. ${\partial_r\Omega_e}$.
Solutions growing with $t$
depend on $\phi_1$.   This growth vanishes if there is no shear.

The main physical consequence of the solution is an algebraic growth on the shortest time scale in the problem, namely the local dynamical time scale $\tau=1/ \Omega_e$. The growth continues at least till the next order terms in $\epsilon$ become important. The exact point of saturation (in this inviscid limit) is under investigation now.
However it is clear that the smaller
$\epsilon=H /R $ is the longer the growth will continue.
The effect is inherent to the thinness of the disk (or equivalently
how cold it is). Also, the solution found here is in the limit of
vanishing viscosity. In reality,
viscous terms would enter at some stage to cause a decay.
%This point is under investigation.
\par
The existence of algebraically growing
solutions is insensitive to the global
entropy gradient, i.e. whether or not it is stable to buoyancy
oscillations, or to the sign of the Rayleigh criterion, i.e. $\partial_r \jmath$.
This may then give the impression that the solution found here
violates the
Solberg-Hoiland criteria (SHc) (e.g. Tassoul, 2000, Rudiger et al., 2002)
which predicts
the instability of linear infinitesimal axisymmetric disturbances
of steady rotating gas flows.  The criteria pertain to
the dynamics of
infintesimal perturbations of the steady solution of the full
Euler equations (\ref{euler_eqns}) and the adiabatic condition (\ref{entropy_eqn})
without any assumption about the size of $\epsilon$.
We refer to this perturbation
procedure as {\em classical-linearization} (CL).
The resulting equations are subsequently analyzed and, following some simplifying assumptions,
imply the SHc.
\par
On the other hand, the approach we have taken here is different
in philosophy: we instead seek to develop a finite amplitude non-linear
solution to the problem of dynamical disturbances of the full equations.
In this approach we start with the apriori assumption that $\epsilon$ is small.
This is followed by assuming that solutions of the equations,
both steady and dynamical,
may be $\epsilon$ expanded.  Solutions are determined by
iteratively solving the resulting equations at each order of $\epsilon$.
Unlike the CL procedure, in which disturbances are infinitesimal,
disturbances here are introduced with amplitudes comparable with
the characteristic parameter of the system, namely $\epsilon$.
It is in this sense that these perturbations are considered finite-amplitude.
\par
Because the two approaches yield different evolution operators
(see below)
it is not surprising that the algebraic growth we discover is
not predicted by the SHc. The SHc are not valid for
finite amplitude perturbations as these but, instead,
they are valid for infinitesimal ones.
%It follows that
%the first set of equations dictating dynamical evolution,
%i.e. (\ref{eq:u}-\ref{eq:P}),
%happen to be linear with an operator that qualitatively
%resembling the one
%resulting from classical-linearization but with certain
%terms missing (see below).  These
%absent expressions lead to the algebraic growth observed in the solution
%calculated here.
%The missing linear terms, together with other effects
%like nonlinear advection, appear at the next order of the expansion procedure.
\par
To put this in somewhat more concrete terms
let
${\bf V}_L=(u_L^{'},\Omega_L^{'},v_L^{'},\rho_L^{'},P_L^{'})^T$,
represent a column vector of the perturbed quantities resulting from the
CL procedure.  In CL
the governing equations for the evolution of the perturbed quantities are,
\beq
\partial_t {\bf V}_L + {\mathbf {\cal P}}{\bf V}_L = 0,\qquad
{\mathbf {\cal P}} =
\left({\matrix{{\bf H}&{\bf \tilde O_e}\cr
{\bf D}&{\bf L} \cr
}}
\right),
\label{big_matrix}
\eeq
with the following operator submatrices defined:
\[
{\bf H} = \left(
\begin{array}{cc}
0 & -2\sfrac{\jmath}{r} \\
\sfrac{1}{r^3}\jmath_r &  0
\end{array}
\right),\quad
{\bf L} = \left(
\begin{array}{ccc}
0 & \sfrac{g_z}{\rho_s} & \sfrac{1}{\rho_s}\partial_z \\
\partial_z\rho_s & 0 & 0 \\
\partial_z P_s + \gamma P_s \partial_z & 0 & 0
\end{array}
\right),
\]
in which $\jmath_r = \partial_r \jmath$,
and,
\[
{\bf D}
=
\left(
\begin{array}{cc}
0 & 0 \\
\sfrac{1}{r}\partial_r  \rho_s r & 0 \\
\partial_r P_s + \gamma P_s\sfrac{1}{r}
\partial_r r & 0
\end{array}
\right),\quad
{\bf \tilde O}_{\bf e}
=
\epsilon^2 \left(
\begin{array}{ccc}
0 & \sfrac{g_r}{\rho_s} &
\sfrac{1}{\rho_s}\partial_r \\
0 & 0 & 0
\end{array}
\right),
\]
with $g_r = -\frac{1}{\rho_s}\partial_r P_s$.  The quantities
$P_s$ and $\rho_s$ are the steady state pressure and
density solutions to (\ref{euler_eqns}) for arbitrary
$\epsilon$.
The SHc refers to the perturbation matrix ${\cal P}$
of the {\em linearized} system.  No reference to the $\epsilon$ parameter is made.
We can go one step further and expand the matrix operator ${\cal P}$ into powers of
$\epsilon$ and when doing so we discover that the $\tilde {\bf O}_{\bf e}$
operator, since it
is proportional to $\epsilon^2$, can be neglected for thin accretion disks (see
below).
\par
By contrast,
having started from (A.1) and expanded in $\epsilon$,
we arrived upon a different system:
one which has taken advantage of the extreme geometry of the system
(i.e. $\epsilon \ll 1$)
but also one which
has  bypassed the assumption of linearization and all of its
host implications.
In particular,
the finite-amplitude
expansion procedure implemented has to lowest order lead to
(\ref{eq:u}-\ref{eq:P}), which is expressed compactly,
$\partial_t {\bf V'} + {\mathbf{{\cal P}_0}}{\bf V'} = 0$,
with
${\bf V}'=(u_1^{'},\Omega_2^{'},v_2^{'},\rho_2^{'},P_2^{'})^T$.
This equation is similar to (\ref{big_matrix}) in which
${\bf V}_\ell \rightarrow {\bf V}'$ and where
${\cal P}_0$ is ${\cal P}$.  The differences between these are
(i) $P_s \rightarrow P_0$ and $\rho_0\rightarrow \rho_s$
and, (ii) ${\bf \tilde O}_{\bf e}$ is replaced by
${\bf O}$, a $2\times3$ zero matrix.  Evidentally ${\cal P}$
and ${\cal P}_0$ differ in quality due to the absence
of ${\bf \tilde O}_{\bf e}$.
\par
The algebraic growth solution discovered
results from this finite-amplitude approach.
The obvious question to ask is:
how do the two approximations relate to each other?
The approximation carried out here is not restricted to
infinitesimal perturbations but instead is one that exploits the geometrical
nature of the object under discussion.
%, i.e. here $\epsilon \ll 1$.
It is therefore clear that the
implications of the SHc analysis and the analysis
presented in this work refer to two disjoint
perturbation spaces which evolve according to different operators.\par
It is easily seen that the work performed by the solution (\ref{uprime_sol}) per
cycle ($2\pi/\Omega_e$) is equal to $2\pi \gamma P_0 \bar u (\partial_r \Omega_e/\Omega_e^2$).
Evidentally this work is zero if the shear vanishes.  It is mainly this extra work
which drives the algebraic growth.
%The integration of the work performed by the solution (\ref{uprime_sol}) over
%a complete cycle does not vanish because of the existence of shear.  It is this
%extra work which drives the growth.

\section{Physical meaning and consequences}
ADs are subject to continuous noise. The consequences of the effect discovered here is a fast rise
in the vertical direction followed, presumably,
by decay.
The process takes place continuously.
The disk is never  'quiet' . The ability to observe such a rise depends on the radial
extent of the initial perturbation. It is clear that the perturbation should be observed to
propagate from small to large radii.
\par
The main result of this work shows that both the vertical velocities and the
density/pressure fluctuations show algebraic growth proportional to $t\sin\Omega_e t$ where
$\Omega_e$ is the disk rotation rate while the radial velocity (and $\dot m$) show no growth  in this approximation - again a result of the staggered nature of the system of asymptotic expansion.
Physically the result holds so long as $t < 1/\epsilon$
for otherwise the asymptotic orderings loses their validity.
Finally the results apply to any rotation law provided the object has $\epsilon \ll 1$.
%The implications to rotating stars where $\epsilon $ is not a good expansion parameter are reported separately.
\begin{acknowledgements}
One of the authors (OMU) would like to thank P. Yecko for insightful discussions
and for calling to our attention the work of Hanifi \& Henningson, (1998).
We would like to thank the anonymous referee for posing questions that helped us to expose
these results with more clarity.  We also thank E.A. Spiegel and O. Regev for insightful
discussions pertaining to these matters.
\end{acknowledgements}

%\vfill
%\eject
%\onecolumn
\appendix
\section{Equations, scalings and expansion}
We  consider the equations of circumstellar flow under the influence of a central
gravitating point source and
subject axisymmetric perturbations. Self gravity is ignored.
The scalings here are identical to those used in the germinal investigations of the these problems,
(e.g. Regev, 1983, Kluzniak \& Kita, 2000, and Regev \& Gitelman, 2002).
The Keplerian
rotation speed  ($\Omega_k$) is set by the Keplerian value at the fiducial radius, i.e.
$\Omega_k^2 \sim GM/\tilde R^3$ with corresponding
Keplerian speed $\tilde V_k^2 \sim GM/\tilde R$.  The gas pressure is given by $P=c^2 \rho$
where $c$ is the speed of sound and the pressure scale is given by  $\tilde P \sim
\tilde \rho \tilde c^2$.\par
Since $H \sim \tilde c^2/g$, it also happens that $\epsilon^2 = \tilde c^2/V_k^2$.
The choice for the temporal scale  is the sound crossing time in the vertical direction, which happens
to be equivalent to the rotation time at the given radius $\tilde R$.
We assume that the radial ($u$) and vertical ($v$) velocities are scaled by the sound speeds and the
rotation rate ($\Omega$)
is scaled by $\Omega_k$.  Substituting  all of the above into the governing equations yields the following,
\beqa
& & \epsilon\partial_t u + \epsilon^2 u \partial_r u
+ \epsilon v\partial_z u - \Omega^2r =
-\epsilon^2\sfrac{1}{\rho} \partial_r P
-\sfrac{1}{r^2}\left[1 + \epsilon^2\sfrac{z^2}{r^2}\right]^{-\sfrac{3}{2}}
\nonumber \label{expanded_u_eqn}\\
& & \rho{\partial_t\Omega}
+ \epsilon\sfrac{\rho u}{r^2}{\partial_r r^2 \Omega}
+\rho v{\partial_z\Omega} = 0, \nonumber \label{expanded_Omega_eqn}\\
& & {\partial_t v} + \epsilon u {\partial_r v}
+v{\partial_z v} = -\sfrac{1}{\rho}{\partial_z P}
- \sfrac{z}{r^3}\left[1 + \epsilon^2\sfrac{z^2}{r^2}\right]^{-\sfrac{3}{2}},\nonumber \\
%&-&\frac{\partial P}{\partial z} - g(r,z)\rho , \nonumber \label{expanded_v_eqn}\\
& & {\partial_t\rho} +
{\epsilon}\sfrac{1}{r}{\partial_r r\rho u}
+ {\partial_z \rho v}  =  0.
 \label{euler_eqns}
 \eeqa
 We assume here short time scale dynamic perturbations so that it is safe to assume adiabatic perturbations,  $dS/dt = 0$, where $S$ is the specific entropy of the gas
 given   by $C_v\ln P/\rho^\gamma$, where $\gamma$ is the usual ratio of specific heats and $C_v$ is the specific heat  at constant volume. Under these assumptions the energy equation becomes:
  \beq
 \left(\partial_t + \epsilon u{\partial_r}
 + v{\partial_z}\right)P + \gamma P
\left(\epsilon \sfrac{1}{r}{\partial_r u}+
{\partial_z v}
 \right) = 0.
\label{entropy_eqn}
 \eeq
 We consider the following asymptotic expansions of the solutions
 \beqa
 \Omega &=& \Omega_0(r) + \epsilon^2\left[\Omega_2(r,z) + \Omega_2'(r,z,t)\right]+ \cdots \nonumber \\
 u &=& \epsilon\left[u_1(r,z) + u_1'(r,z,t)\right] + \cdots \nonumber \\
 v &=& \epsilon^2\left[v_2(r,z) + v_2'(r,z,t)\right] + \cdots \nonumber \\
P &=& P_0(r,z) + \epsilon^2\left[P_2(r,z) + P_2'(r,z,t)\right]+ \cdots \nonumber \\
\rho&=& \rho_0(r,z) + \epsilon^2\left[\rho_2(r,z) + \rho_2'(r,z,t)\right]+ \cdots
 \eeqa
To lowest order we find the standard thin disk solutions which
do not depend at all on any viscous mechanism:
\begin{equation}
\Omega_0 = r^{-{3}/{2}}, \quad \partial_z P_0 = -\rho_0 g(r,z),
\quad g(r,z) \equiv z/r^3,
\end{equation}
together with barotropic equation of state.  The equations governing
the dynamics of the temporally evolving quantities (i.e. $\rho_2',P_2',v_2',u_1',\Omega_2'$) are quoted in
the body of the text.
The solutions to the steady quantities $\rho_2,P_2,v_2,u_1,\Omega_2$ are
not sought here but are detailed in Umurhan et al (2005, in preparation).  We do note that in the absence of
viscosity   $v_i$ and $u_i$ vanish to all orders  unless we externally impose some amount
of radial mass flux into the system.
%Furthermore, if the basic
%state relationship is barotropic then the correction to the steady rotation speed
%will be independent of the coordinate $z$ - which is
%a restatement of the well known fact that rotating barotropes have azimuthal velocity
%profiles which are constant on cylinders (Tassoul, 2000).
\end{document}